# Searching for Large-gap Quantum Spin Hall Insulators: Boron-Nitride/(Pb, Sn)/α-Al$_2$O$_3$ Sandwich Structures

Hui Wang[1,3,+], D. Lu[1,2,+], J. Kim[1], Z. Wang[3], S. T. Pi[1] and R. Q. Wu[1,3,*]

[1]*Department of Physics and Astronomy, University of California, Irvine, CA 92697-4575, USA*

[2]*Department of Physics, Nanjing University Aeronautics and Astronautics, Nanjing 211106, CHINA*

[3]*State Key Laboratory of Surface Physics and Department of Physics, Fudan University, Shanghai 200433, CHINA*

Topological insulators hold great potential for efficient information processing and storage. Using density functional theory calculations, we predict that a honeycomb lead monolayer can be stabilized on the Al$_2$O$_3$ (0001) substrate and becomes topologically non-trivial with a sizeable band gap (~0.27 eV). Furthermore, we propose to use hexagonal boron-nitride (h-BN) monolayer as a protection for the topological states of Pb/Al$_2$O$_3$ and Sn/Al$_2$O$_3$. Our findings suggest new possibilities for designing and protecting two-dimensional TIs for practical applications.

The topological state of matter is one of the most attractive interdisciplinary research subjects in condensed matter physics and material science [1-4]. In particular, topological insulators (TIs) featured with their topologically protected edge states inside bulk band gaps are of great potential for application in coherent non-dissipative spintronic devices [5-7]. Two-dimensional (2D) TIs have received particular research interests because they manifest the peculiar quantum spin Hall (QSH) effect and have several intrinsic advantages such as structural simplicity as well as the easiness for characterization and manipulation. The two well-established theoretical models for achieving the topological state of 2D materials are 1) the Kane-Mele (KM) type that is based on the special band feature of honeycomb lattices such as graphene, [4] and 2) the Bernevig-Hughes-Zhang (BHZ) type that is based on the band inversion of states in quantum wells [2]. So far, the QSH phase has only be observed in very few systems such as HgTe/CdTe [8] and InAs/GaSb [9] quantum wells, whereas no experimental confirmation for the KM type 2D TIs has been reported after tremendous efforts in the past few years [10-17]. The search for new topological materials [18, 19] and the realization of the QSH effect in 2D honeycomb lattices are still main challenges in this realm.

One of the main hurdles for the construction of excellent 2D TIs is to have both large bulk gap and high structural stability. Unlike traditional semiconductors, band gaps of TIs result from the spin orbit coupling (SOC) effect that is extremely small for elements with stable honeycomb structures (e.g., carbon). To enhance the SOC in 2D materials, many approaches have been proposed, from incorporating heavy adatoms [11], using chemical functionalization [13, 20, 21] to choosing appropriate substrates [12, 22-26]. However, various unfavorable factors, such as clustering of adatoms, unachievable decoration configurations, and inherent structural instability, impose challenges for the experimental realization of these proposals. Recently, honeycomb monolayers of other group-IV elements such as Si, Ge and Sn have received particular attention due to the possibility of having intrinsically large band gaps (up to 0.3 eV) [10, 13, 27]. Although some of them have been successfully synthesized on different substrates, the utilization of their Dirac states is impeditive due to detriments such as structural instability even in lab conditions [28-32]. For the further development of these materials, one needs to (1) find a benign substrate that provides stable mechanical support yet retains the peculiar topological properties of potential 2D TIs; (2) prevent the aging effect of chemically active 2D TIs in ambient conditions.

In this paper, we report results of systematic density functional theory (DFT) calculations and tight binding (TB) modeling for the topological features of BN/(Pb, Sn)/$Al_2O_3$ (0001) sandwich structures. Here we demonstrate that the $Al_2O_3$ (0001) substrate can provide stable support for the honeycomb Pb and Sn monolayers with large binding energies. Remarkably, addition of a hexagonal BN monolayer on (Pb,Sn)/$Al_2O_3$ (0001) retains the topological features and hence the QSH state of BN/(Pb, Sn)/$Al_2O_3$ (0001) is robust and well-protected, suitable for applications in the ambient condition. Furthermore, it is interesting that the hybridization between Pb and $Al_2O_3$ (0001) converts the topologically trivial Pb monolayer to a 2D TI, according to analyses in the $Z_2$ invariants and the robust edge states. One of the main goals in this realm is searching for feasible ways to fabricate stable functional 2D TIs, our studies point out a new route toward that direction, by choosing appropriate adlayer and substrate to generate and protect the topological phase of heavy elements.

DFT calculations were performed with the Vienna Ab-initio Simulation Package (VASP) [33] at the level of spin-polarized generalized gradient approximation (GGA) [34]. We treated O-2$s$2$p$, B-2$s$2$p$, N-2$s$2$p$, C-2$s$2$p$, Al-3$s$3$p$, Sn-5$s$5$p$ and Pb-6$s$6$p$ as valence states and adopted the projector-augmented wave (PAW) pseudopotentials to represent the effect of their ionic cores [35, 36]. The spin-orbit coupling term was incorporated self-consistently using the non-collinear mode [37, 38]. To obtain reliable adsorption geometries and binding energies, the nonlocal van der Waals functionals (optB86b-vdW) were also included [39-42]. The energy cutoff for the plane-wave expansion was set to as high as 700 eV, sufficient for this system according to our test calculations. A 9×9×1 Monkhorst-Pack k-point mesh was used to sample the Brillouin zone [43]. Positions of all atoms except those in the central six $Al_2O_3$ layers (out of the eighteen $Al_2O_3$ layer slab for the substrate) were optimized with a criterion that the atomic force on each atom becomes smaller than 0.01 eV/Å and the energy convergence is better than $10^{-6}$ eV. In the following, we use Pb/$Al_2O_3$ as the example to illustrate the topological features of Pb and Sn monolayers as well as the protective role of a hexagonal BN layer (h-BN) on top of them.

Lead is the last and heaviest group IVB element. Comparing to its siblings (C, Si, Ge, Sn), lead monolayer deserves a particular attention due to its strongest SOC. As shown in the inset of Fig. 1(a), buckled honeycomb lead monolayer (denoted as "h-Pb") with an optimized lattice constant a=4.921 Å (Pb-Pb bond length d=2.99 Å) is energetically more stable (0.4 eV/unit cell lower) than the planar case (see supplemental material Fig. S1). Noticeably, both lead and tin monolayers may keep the stable honeycomb geometry with chemical decoration (denoted as "h-Pb-X" and "h-Sn-X", X = H, F, Cl, Br, I and At) [13], as depicted in the inset of Fig. 1(b).

The band structures of the pristine ("h-Pb") and hydrogen decorated lead monolayer ("h-Pb-H") are shown in Fig. 1(a) and (b). For "h-Pb", two pairs of bands touch at $\Gamma$ and K points in the absence of SOC, corresponding to the $p_{xy}$ and $p_z$ orbitals as seen from the projected density of states (PDOS) in Fig. 1(c). In "h-Pb-H", only one pair bands touch at the $\Gamma$ point. This can be explained by the hybridization between H-$s$ and Pb-$p_z$ orbitals near the Fermi energy [see Fig. 1(c)], giving rise to a large band gap at the K point as shown in Fig. 1(b). When SOC is turned on, a global gap is opened and they become 2D insulators. To determine if the band gaps are topologically nontrivial, the k-dependent berry curvatures of "h-Pb" and "h-Pb-H" are calculated and the results are given in the insets of Fig. 1(a) and (b). Note that the topological nature can be determined from $Z_2$ invariants that rely on matrix elements of Bloch wave functions at time-reversal-invariant-momentum (TRIM) points in the Brillion Zone (BZ) [44]. Here, the band topology are evaluated through both band parities [45] at the $\Gamma$ and M points and the n-field method [46]. Interestingly, pristine lead monolayer "h-Pb" is a trivial insulator (band gap: ~0.45 eV) while "h-Pb-H" is topological insulator (band gap: ~0.96 eV) [see Fig. 1(b)].

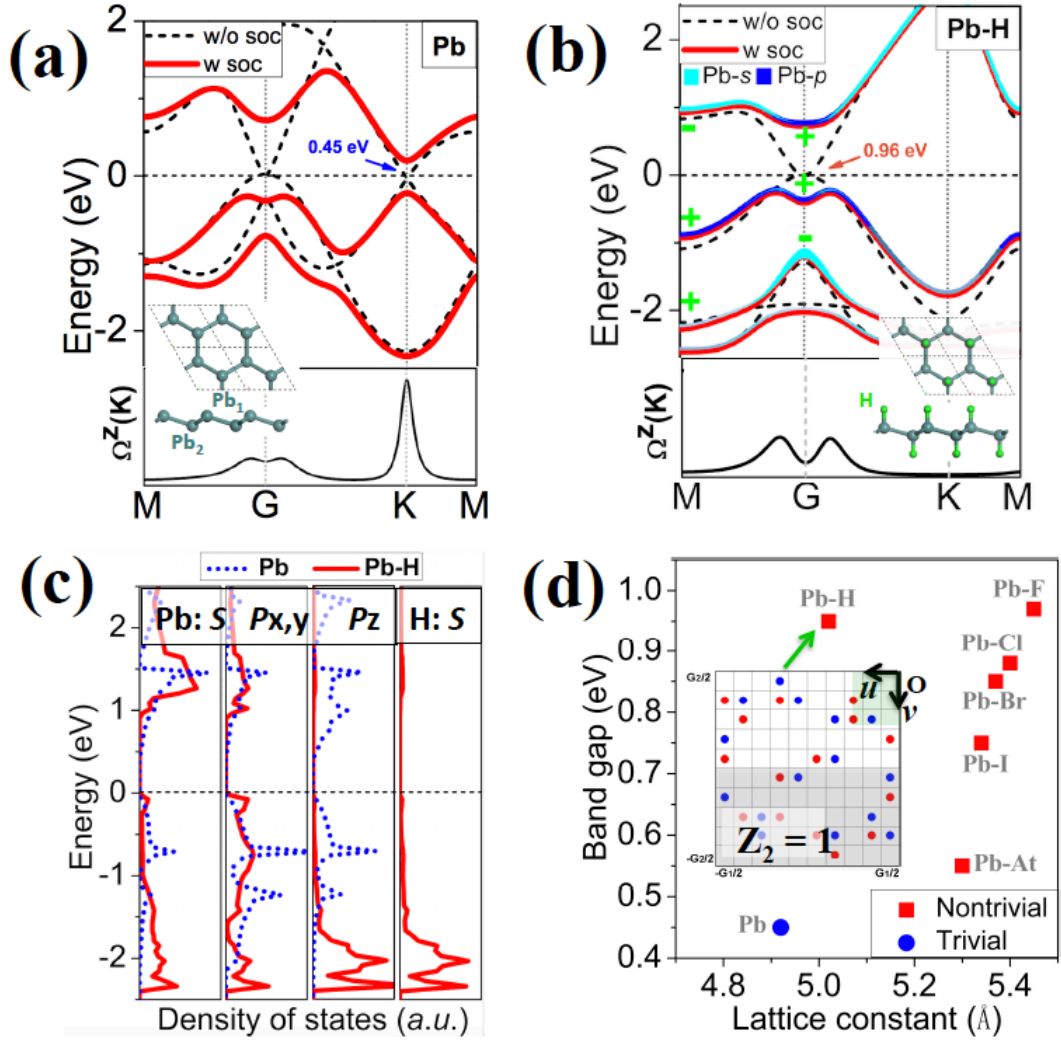

Fig. 1 (color online) (color online) Electronic band structures of (a) Pb and (b) hydrogen-decorated Pb (Pb-H) monolayer with (thick red solid lines) and without (thin black dash line) spin orbit coupling in calculations, light green and blue in (b) indicated the weights of s and p orbital of Pb atoms. Fermi level is set at 0, band parities are marked at TRIM points. Insets in (a) and (b) are the corresponding atomic structures and spin berry curvature along the momentum path of band structures. (c) The atomic-orbital projected local density of states of Pb and Pb-H layer, Pb and H atoms are demonstrated for the $p_z$, $p_x+p_y$ and s orbitals, respectively. (d) Phase diagram for band topologies and lattice of Pb-X. Inset is the n-field configuration of Pb-X. The calculated torus in Brillouin zone is spanned by $G_1$ and $G_2$. Note that the two reciprocal lattice vectors u and v actually form an angle of 120°. The red and blue circles denote n = 1 and

*n = -1, respectively, while the blank denotes n=0. The $Z_2$ invariant is 1, by summing the n-field over half of the torus, as shown in the shaded yellow area.*

Since chemical decoration plays a pivotal role in tuning the QSH state of "h-Pb", we also calculated the lattice constants, band gaps and topological properties of "h-Pb" with other different functional groups such as: -F, -Cl, -Br, -I and -At. As shown in Fig. 1(d), all these functional groups increase the in-plane lattice constant of "Pb" by up to ~10%. More importantly, they all lead to topological nature with nontrivial band gaps in a range of 0.55~1.0 eV. However, the structural stability of 2D monolayers is a critical issue for their fabrication. We conducted *ab initio* molecular dynamic (AIMD) simulations for "h-Pb" and "h-Pb-H" at room temperature (300K) and found that they are dynamically unstable as seen from the severe oscillation of total energy after a few picoseconds in Fig. S2. The snapshots of the final atomic structure in AIMD clearly demonstrate that "h-Pb" and "h-Pb-H" are completely damaged from their initial structures (see insets in Fig. S2 (c) and (d)). Obviously, one needs a good substrate to ensure the structural stability of "h-Pb" and "h-Pb-X".

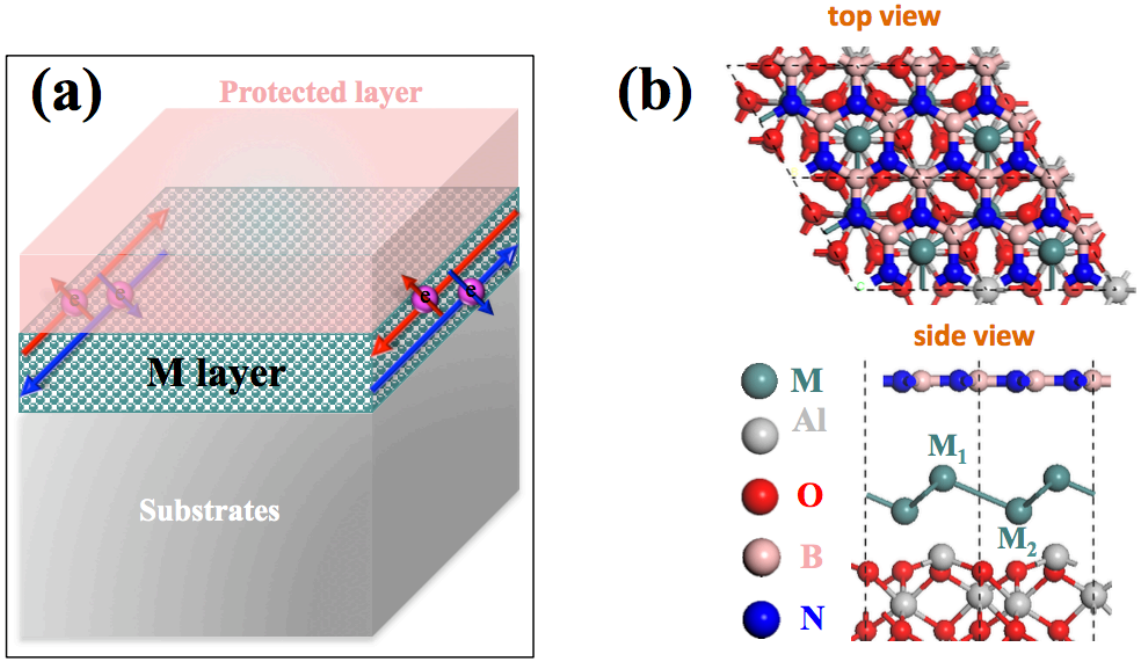

*Fig. 2 (color online) (a) Schematic model for freestanding M layer adsorbed on substrate and covered by a protected layer, where M are Pb and Sn atoms, respectively. (b) Proposed sandwiched structures of BN/(Pb,Sn)/Al$_2$O$_3$, Al and oxygen atoms are depicted by grey and red balls. where boron and nitrogen atoms are represented by pink and blue balls, respectively.*

As sketched in Fig. 2(a), we may place "h-Pb" on substrates with small lattice mismatch such as Si$_3$N$_4$ (0001) [47], GaAs (111) [48] and Al$_2$O$_3$ (0001) [49, 50]. Here, we report that Al$_2$O$_3$ (0001) is a good candidate for the realization of the 2D TI state of Pb monolayer. As shown in Fig. 2(b), Pb$_1$ atoms tightly bind on atop-Al with a binding energy of ~ -0.78 eV per Pb atom. "h-Pb" on Al$_2$O$_3$ maintains its buckled structure with a slight expansion in the Pb-Pb distance, to 3.06 Å. Similar to Sn/Al$_2$O$_3$ [12], Al$_2$O$_3$ substrate induces a noticeable charge redistribution in "h-Pb", with a charge transfer of 0.2 e from the $p_z$ orbital of Pb$_1$ to Pb$_2$, according to Bader charge analysis [51]. As shown in Fig. 3(a), most bands around the Fermi level are from Pb, as depicted by the red curves. When SOC is excluded in our DFT calculations, Pb/Al$_2$O$_3$ behaves like a poor metal with two bands touching the Fermi level in the vicinity around the $\Gamma$ point. One can also see that band touching of "Pb" at K points is removed in Pb/Al$_2$O$_3$, indicating the similar effect of substrate toward modifying the electronic states of "h-Pb" as does the X-decoration.

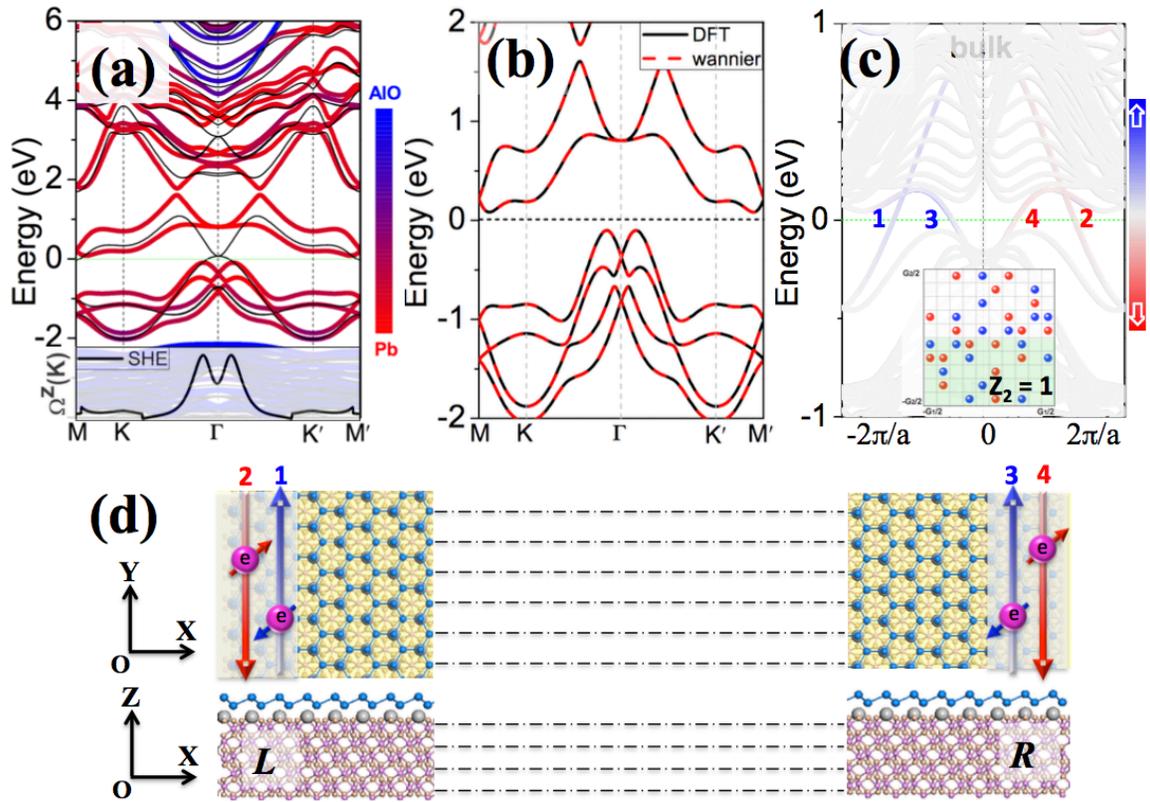

Fig. 3 (color online) (a) Electronic bands of Pb/Al$_2$O$_3$ with (thick colored lines) and without (thin black line) spin orbit coupling in calculations. The color bar indicates the weights of Pb (red) and Al$_2$O$_3$ substrate (blue). Green dashed line represents the Fermi level. Inset is the corresponding spin berry curvature along the momentum path of band structures. (b) Comparison of band structures for Pb/Al$_2$O$_3$ obtained from DFT (black solid line) and Wannier function (red dash line) method. (c) Spin projections of helical edge states of Pb/Al$_2$O$_3$ nanoribbon from wannier TB modeling. Blue and red represent spin up and spin down bands, respectively. Band 1, 2 and 3, 4 demonstrate the left and right edge states corresponding

*to the sketched atomic models in (d). The n-field over half of the torus is depicted in the inset. (d) Top and side views for the schematic model of Pb/$Al_2O_3$ nanoribbon along the y direction, the spin current protected by the topological nature are demonstrated at the edge of nanoribbon. Symbols "L" and "R" in side view represent the left and right side of nanoribbon, respectively.*

Now, the most important question is that whether Pb/$Al_2O_3$ is a topological insulator. To answer this question, we include SOC effects in the calculation and a gap as large as 0.27 eV opens as shown in Fig. 3(a), indicative of the nontrivial topological feature of Pb/$Al_2O_3$. Moreover, diverge spin berry curvature can be found near the $\Gamma$ point, due to the SOC inverted Bloch wave function as depicted in the inset of Fig. 3(a). Further n-field method calculation at TRIM points confirms the topological nature of Pb/$Al_2O_3$, as shown in the inset of Fig. 3(b). As another remarkable feature of 2D TIs, we also examined the existence of gapless edge states with the spin-momentum locking feature in the bulk band gap for a Pb/$Al_2O_3$ nanoribbon. Since the main features near the Fermi level stem from the *p* orbitals of Pb atoms (shown by PDOS in Fig. 2(c)), we used the maximally localized Wannier functions (MLWFs) to interpolate these bands. [52] The quality of Wannierized band is high in Fig. 3(b). This allows us to construct the tight-binding (TB) Hamiltonian of zigzag nanoribbons of Pb/$Al_2O_3$ with a large width ($d \approx 20$ nm) as sketched in Fig. 3(d). As shown in Fig. 3(c), the calculated band structure of the zigzag nanoribbon clearly show several states within the bulk gap (grey lines). We further made spatial and spin projections of these states and presented them with red and blue in Fig. 3(c). One can see the spin-momentum locking feature of 1D helical electrons (left edge) and holes (right edge) for Pb/$Al_2O_3$ nanoribbon, an unambiguous characteristic of the QSH state. These edge states have opposite group velocities at opposite edges, in agreement with previous reports [27, 53]. The existence of gapless helical edge states and non-zero $Z_2$ topological invariants clearly show that the Pb/Al$_2$O$_3$ is a QSH insulator. One may note that "h-Pb-H" can also bind to Al$_2$O$_3$ through van der Waals forces, leaving the main feature of its band and topological properties unchanged as shown in Fig. S3.

TABLE 1. The binding energy ($E_b$) of BN, distance between Al and atop-M ($d_{Al-M}$), BN and atop-M ($d_{BN-M}$), and bond length ($d_{M-M}$) for the most stable BN/M/Al$_2$O$_3$ sandwiched structures, where M represent Pb and Sn atoms, respectively.

|            | $E_b$ (eV) | $d_{Al-M}$ (Å) | $d_{BN-M}$ (Å) | $d_{M-M}$ (Å) |
|------------|------------|----------------|----------------|---------------|
| BN/Pb/AlO  | -0.29      | 3.07           | 3.54           | 3.05          |
| BN/Sn/AlO  | -0.25      | 2.88           | 3.56           | 3.03          |

Owing to the high chemical activity of Pb atoms in Pb/Al$_2$O$_3$ (Sn atoms Sn/Al$_2$O$_3$ in previous studies [12]), appropriate protection of their electronic structures is another critical issue for practical device applications. To this end, we consider placing a layer of 2D materials with high structural stability, such as graphene (Gr), boron nitride (h-BN), etc. According to our calculations, the Dirac states of graphene may exist in the band gap of Pb/Al$_2$O$_3$ and Sn/Al$_2$O$_3$ as shown in Fig. S4 and hence Gr/(Pb,Sn)/Al$_2$O$_3$ are poor metals. In contrast, h-BN monolayer has a large band gap ~5 eV. In Table S1, Table S2 and Table 1, one may find that binding energies of h-BN monolayer on Pb/Al$_2$O$_3$ and Sn/Al$_2$O$_3$ are about -0.29 eV and -0.25 eV, and the geometry of Pb/Al$_2$O$_3$ and Sn/Al$_2$O$_3$ is hardly affected by the presence of h-BN. The band structures of h-BN/(Pb,Sn)/ in Fig. 4(a) and (c) indicate that the low energy band features of (Pb,Sn)/Al$_2$O$_3$ are almost unchanged. Further analysis of the PDOS of h-BN/(Pb, Sn)/Al$_2$O$_3$ (Fig. 4(b) and (d)) shows that p-orbitals of B and N atoms are either below -1.5 eV or above 3.5 eV with respect to the Fermi level. As a result, electronic states near the Fermi level are very similar to those of (Pb,Sn)/Al$_2$O$_3$. Therefore, h-BN monolayer is a good candidate for protecting the topological states and transport properties of (Pb,Sn)/Al$_2$O$_3$.

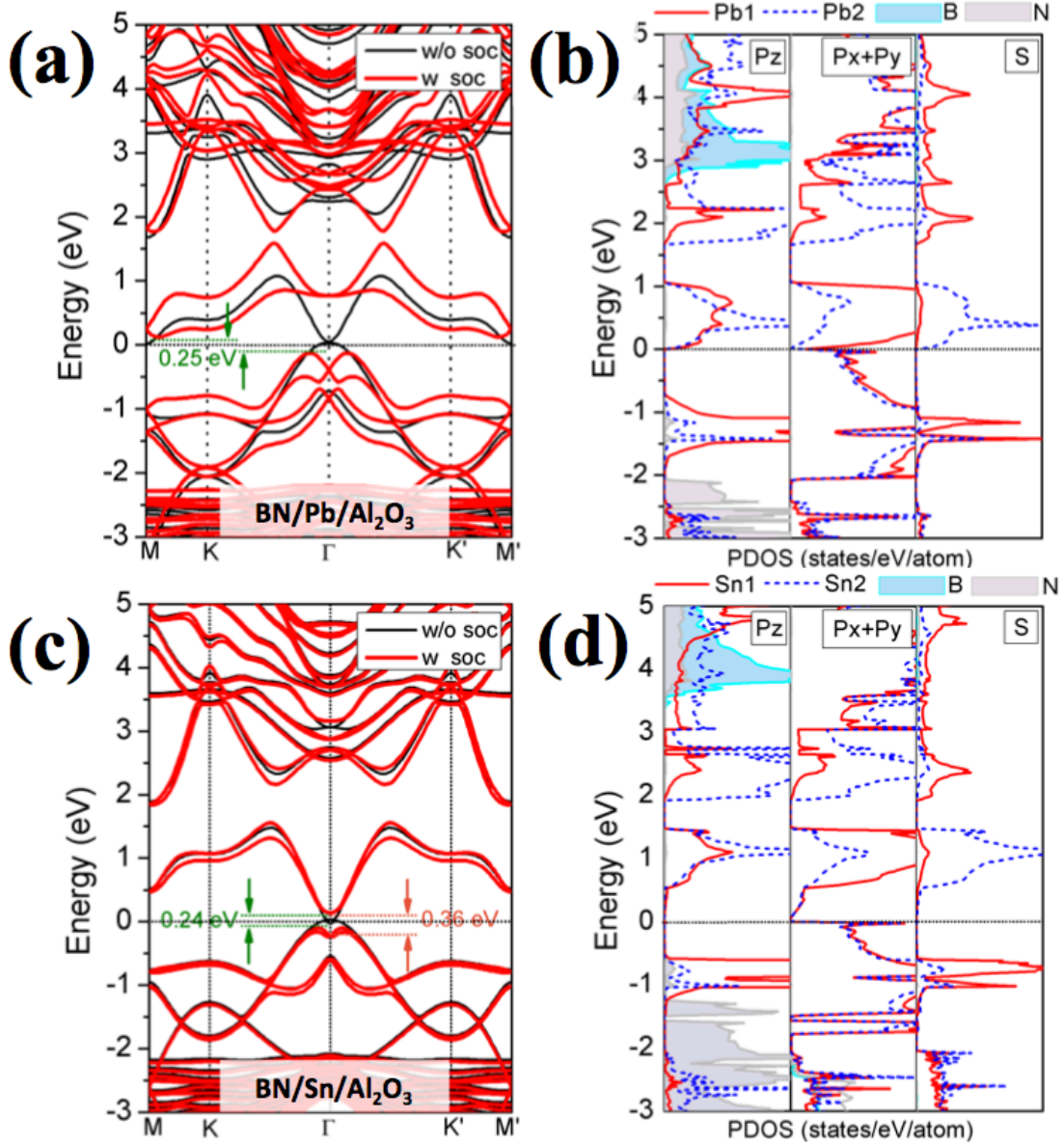

Fig. 4 (color online) Electronic band structures of (a) BN/Pb/Al$_2$O$_3$ and (c) BN/Sn/Al$_2$O$_3$ sandwiched structures with (thick red solid lines) and without (thin black line) spin orbit coupling in calculations. (b)

*and (d) are the corresponding atomic-orbital projected local density of states of B, N, Pb and Sn atoms, where B, N, Pb and Sn atoms are demonstrated for the $p_z$, $p_x+p_y$ and s orbitals, respectively.*

In summary, we have shown that $Al_2O_3$ (0001) is an excellent substrate for both stabilizing "h-Pb" and "h-Sn" for the realization of the QSH effect in 2D materials. Furthermore, it may also tune the topologic phase of "h-Pb" from trivial to nontrivial, as confirmed by the direct calculations of the $Z_2$ topological invariant and gapless edge states. As a step toward applications, we propose to use h-BN monolayer for protecting the QSH states of (Pb,Sn)/$Al_2O_3$. Since Pb is a good superconductor, one may add more Pb layers on BN/Pb/$Al_2O_3$ to create a TI/superconductor combination with the same element for studies of Majorana fermions. The present studies point out a new way of designing 2D TIs with large band gap and high stability through an unconventional means: hybridization with insulating substrates and protected by insulating monolayers.

Work at UCI was supported by DOE-BES (Grant No: DE-FG02-05ER46237) and SHINES (Grant No. SC0012670). Work at Fudan was supported by the CNSF (Grant No: 11474056) and NBRPC (Grant No: 2015CB921400). Computer simulations were partially supported by National Energy Research Scientific Computing Center (NERSC).

* To whom correspondence should be addressed (wur@uci.edu)
+ These authors contributed equally to this work.

Supplementary materials on:

# Searching for Large-gap Quantum Spin Hall Insulators: Boron-Nitride/(Pb, Sn)/α-Al$_2$O$_3$ Sandwich Structures

Hui Wang[1,3,+], D. Lu[1,2,+], J. Kim[1], Z. Wang[3], S. T. Pi[1] and R. Q. Wu[1,3,*]

[1]*Department of Physics and Astronomy, University of California, Irvine, CA 92697-4575, USA*

[2]*Department of Physics, Nanjing University Aeronautics and Astronautics, Nanjing 211106, CHINA*

[3]*State Key Laboratory of Surface Physics and Department of Physics, Fudan University, Shanghai 200433, CHINA*

TABLE S1. The binding energy ($E_b$) of BN, distance between Al and atop-Pb ($d_{Al-Pb}$), BN and atop Pb ($d_{BN-Pb}$), and bond length ($d_{Pb-Pb}$) at different adsorption sites, respectively.

| BN | $E_b$ (eV) | $d_{Al-Pb}$ (Å) | $d_{BN-Pb}$ (Å) | $d_{Pb-Pb}$ (Å) |
|---|---|---|---|---|
| T | -0.28 | 3.07 | 3.61 | 3.05 |
| H | -0.29 | 3.07 | 3.53 | 3.05 |
| B | -0.28 | 3.07 | 3.57 | 3.05 |

TABLE S2. The binding energy ($E_b$) of BN, distance between Al and atop-Sn ($d_{Al-Sn}$), BN and atop-Sn ($d_{BN-Sn}$), and bond length ($d_{Sn-Sn}$) at different adsorption sites, respectively.

| BN | $E_b$ (eV) | $d_{Al-Sn}$ (Å) | $d_{BN-Sn}$ (Å) | $d_{Sn-Sn}$ (Å) |
|---|---|---|---|---|
| T | -0.24 | 2.87 | 3.64 | 3.03 |
| H | -0.25 | 2.88 | 3.55 | 3.03 |
| B | -0.24 | 2.88 | 3.58 | 3.03 |

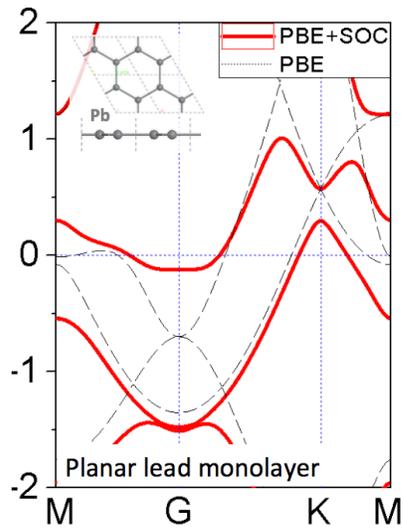

Fig S1 (color online) Electronic band structures planar lead monolayer with (red thick line) and without (back thin line) SOC. Insets demonstrate the top and side view of atomic structures with a lattice constant of 5.159 Å. Horizontal blue-dashed line indicates the Fermi level.

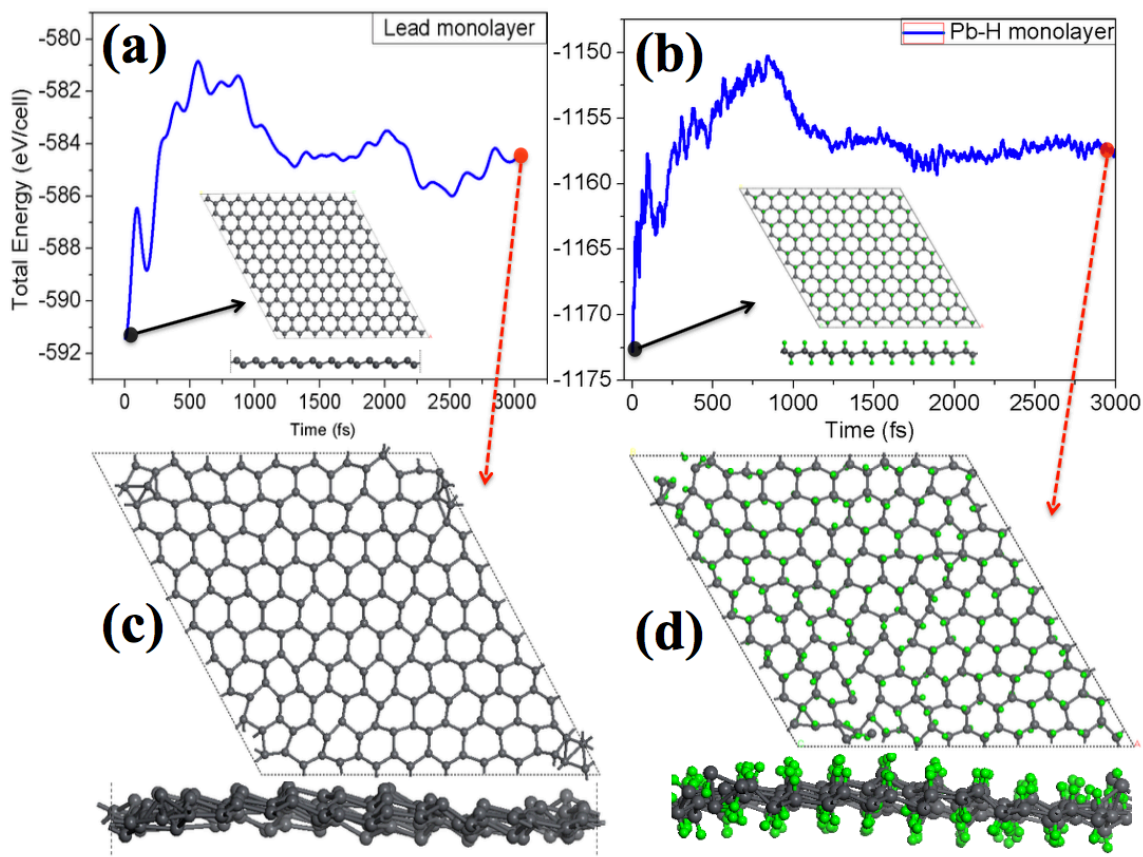

Fig S2 (color onine) Ab initio molecular dynamic simulation of monolayer (a) "Pb" and (b) "Pb-H" with a period of 3 pico seconds. Insets demonstrate the initial atomic structures. (c) and (d) represent snapshots of corresponding to the final structures obtained from AIMD.

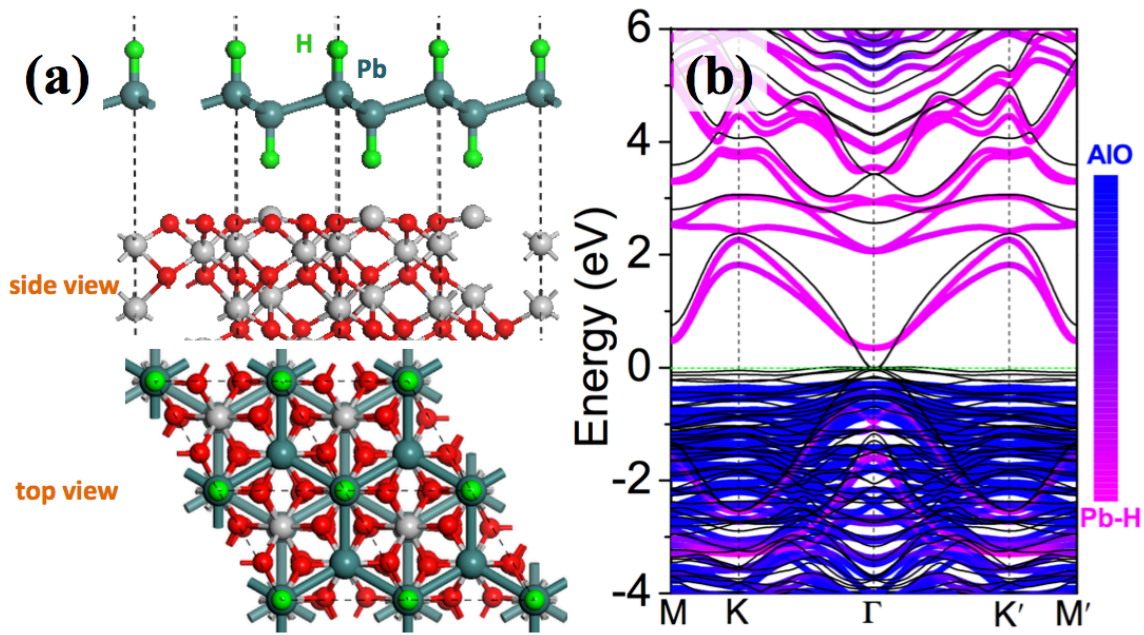

Fig S3 (color onine) (a) Side and top view for the atomic structures of Pb-H/$Al_2O_3$. (b) Electronic band structures of Pb-H monolayer adsorbed on α-alumina surface with (color thick line) and without (back thin line) SOC. The color bar indicates the contributions from Pb-H layers (pink) and α-alumina substrate (blue), with the atomic contributions computed by projecting the Bloch wave functions of Pb-H/$Al_2O_3$ into Pb-H atoms. Horizontal green-dashed line indicates the Fermi level.

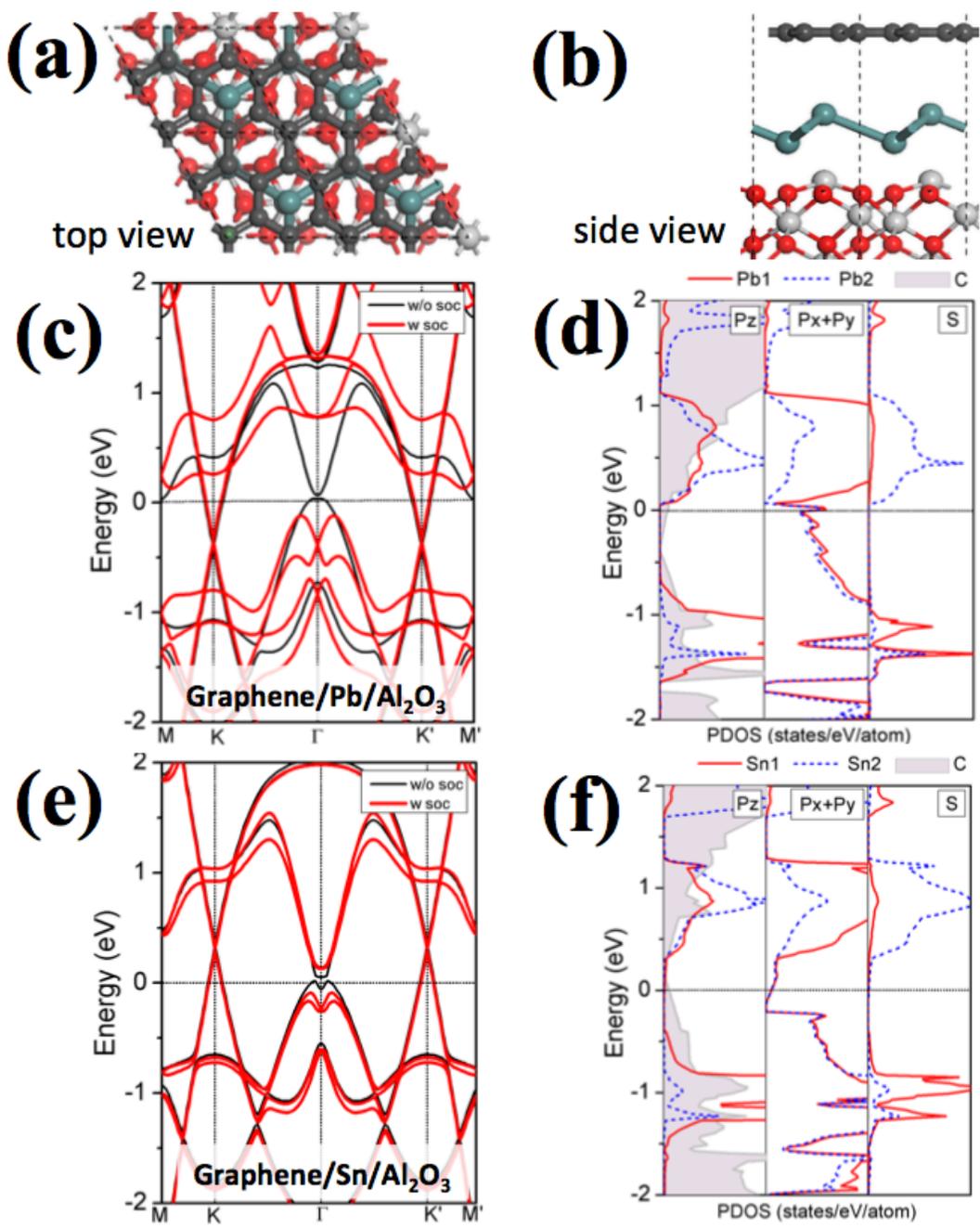

Fig. S4 (color online) (a) Top and (b) side view of sandwiched graphene/(Pb, Sn)$Al_2O_3$ structures. Electronic band structures of (c) graphene/Pb/ $Al_2O_3$ and (e) graphene/Sn/$Al_2O_3$ sandwiched structures with (thick red solid lines) and without (thin black line) spin orbit coupling in calculations. (d) and (f) are the corresponding atomic-orbital projected local density of states of C, Pb and Sn atoms, where C, Pb and Sn atoms are demonstrated for the $p_z$, $p_x+p_y$ and s orbitals, respectively.